\documentclass[printer]{tJOT2e}
\usepackage{rotating}
\usepackage{amsmath}
\usepackage{graphicx,epsfig,floatflt,subfigure}
\newcommand\ie{{\it i.e.}}

\begin{document}

\doi{10.1080/14685240xxxxxxxxxxxxx}
\issn{1468-5248}
%\issnp{} \jvol{00} \jnum{00} \jyear{2008} \jmonth{September}

\markboth{W.J.T. Bos and J.-P. Bertoglio}{Lagrangian Markovianized Field Approximation for turbulence}

\title{Lagrangian Markovianized Field Approximation for turbulence}

\author{Wouter J.T. Bos$^\ast$\thanks{$^\ast$ wouter.bos@ec-lyon.fr},
and Jean-Pierre Bertoglio\\
\vspace{0.2cm}
LMFA-CNRS, Universit\'e de Lyon, Ecole Centrale de Lyon, 69134 Ecully, France}
\received{2012}

\maketitle

\begin{abstract}
In a previous communication (W.J.T. Bos and J.-P. Bertoglio 2006, Phys. Fluids , 18, 031706), a self-consistent Markovian triadic closure was presented. The detailed derivation of this closure is given here, relating it to the Direct Interaction Approximation and Quasi-Normal types of closure. The time-scale needed to obtain a self-consistent closure for both the energy spectrum and the scalar variance spectrum is determined by evaluating the correlation between the velocity and an advected displacement vector-field. The relation between this latter correlation and the velocity-scalar correlation is stressed, suggesting a simplified model of the latter. The resulting closed equations are numerically integrated and results for the energy spectrum, scalar fluctuation spectrum and velocity-displacement correlation spectrum are presented for low, unity and high values of the Schmidt number.
\end{abstract}

\bigskip

%\pacs{47.27.Ak, 47.27.Eq, 47.27.Gs, 47.27.Jv, 47.27.Te, 47.27.Qb}

%\pacs{47.27.Ak, 47.27.Eq, 47.27.Gs, 47.27.Jv, 47.27.Te, 47.27.Qb}

\section{Introduction}

The stochastic nature of turbulence introduces the necessity of a statistical description. Typical statistical descriptors are the mean velocity and the moments of the turbulent fluctuations. In isotropic turbulence the mean velocity is zero and the velocity fluctuations reduce to their geometrically simplest form. Still the interaction of a large number of lengthscales renders the description difficult. A phenomenological description of isotropic turbulence was proposed by Kolmogorov \cite{Kolmogorov}, leading to a prediction of the distribution of energy over different lengthscales in a turbulent flow assuming that high Reynolds number turbulence is locally isotropic and that mode interactions are local in scale-space. Experiments \cite{Grant1962}, presented at the famous Marseille Conference in 1961, confirmed the predicted energy distribution, proportional to $k^{-5/3}$, with $k$ the wavenumber. Deriving this description from first principles, \emph{i.e.} starting from the Navier-Stokes equations, is 
a formidable task. Indeed phenomenological models (\emph{e.g.} references \cite{Leith,Heisenberg}) reproduced the correct wavenumber dependence of the energy distribution, but the models did not have any direct relation to the Navier-Stokes equations.

A first attempt to build a model for the multi-scale dynamics of a turbulent flow could start from the observation that the single-point velocity statistics seem close to Gaussian. If is assumed that all velocity moments behave in a Gaussian manner, one is directly led to the erroneous conclusion that all different lengthscales are independently decaying under the influence of viscosity since the interaction between modes is governed by triple correlations which are zero in a Gaussian field. Indeed, in a Gaussian field modes are statistically independent and this precludes energy transfer among scales. It is thus clear that at the level of third order moments, cumulants, defined as the differences of quantities with respect to their Gaussian values, can not be neglected. 

A logical attempt to obtain a correct description of turbulence would then consist in closing the hierarchy of moments by an assumption of Gaussianity at the first level which does allow the triple correlations to be non-zero, in other words, at the level of quadruple correlations. This quasi-normal assumption \cite{Millionschikov1941,Tatsumi1957,Proudman1954} is well documented in the textbook by Monin and Yaglom \cite{Monin}. Unfortunately, this approach leads to non-physical results, since the energy-spectrum predicted by the quasi-normal model becomes negative at long-times \cite{Ogura1963}, as was carefully suggested previously by Kraichnan \cite{Kraichnan1957}. 

A great breakthrough in the theory of turbulence was the introduction of the Direct Interaction Approximation (DIA) \cite{KraichnanDIA}. The great difference with the foregoing attempts is that it is a two-time theory, introducing a new quantity, the response function, corresponding to the response of a turbulent flow on an infinitesimal perturbation.
The big difference with the quasi-normal approximation is that cumulants of all orders are allowed to be non-zero and quadruple moments are not assumed to be Gaussian. This issue was at the heart of the discussion between Ian Proudman and Robert Kraichnan at the conference in Marseille \cite{Proudman1962,Kraichnan1962} (see also \cite{Rubinstein2012}).
Later works showed how these higher order cumulants can be determined within the DIA framework \cite{Chen1989,Bos2012-3}. The DIA theory suffered however from both a theoretical and a practical weakness. 

The theoretical weakness consists in the fact that DIA does not predict the correct inertial range behavior (\emph{i.e.} in agreement with Kolmogorov's prediction). This was traced back to the fact that its dynamical behavior is not invariant to Galilean transformations \cite{Kraichnan1964} and the defect was corrected for by reformulating the theory in a Lagrangian coordinate system \cite{Kraichnan65}. This yielded a description, called the Lagrangian History DIA which gives correct asymptotic behavior for the energy spectrum. However, the formulation becomes extremely complicated in this new set of coordinates. A simplified, or abridged, version was proposed in the same work and an even simpler description of the same type is the Lagrangian Renormalized Approximation (LRA) \cite{Kaneda81,Kaneda1986}.

The practical weakness of these theories, both DIA and the Lagrangian versions, is the two-time character which precludes a computation of long-time statistics, unless the time-history is truncated at some time difference. To obtain a practical statistical model for the dynamics of turbulence, one needs a procedure called Markovianization. This procedure is based on the assumption that the statistical two-time behavior of the turbulent quantities is known, so that the theories can be simplified to a dynamical description which only depends on the current time. Kraichnan proposed such a procedure and applied it to DIA, yielding an approximation known as the Test Field Model \cite{KraichnanTFM}. The two-time quantities are in this approach assumed to be exponentially decaying with a typical time-scale which is determined by comparing the decorrelation of a fully compressible field, the test field, with the decorrelation of the advecting velocity field. A simplified version of the Test Field Model is the Eddy-
Damped Quasi-Normal Markovian model \cite{Orszag}, in which this time-scale is heuristically modeled by a phenomenological straining time-scale of the Heisenberg type \cite{Heisenberg}.

In a previous work \cite{Bos2006} we proposed a method to determine this time-scale self-consistently in the framework of the EDQNM theory, using the observation that this time-scale in Lagrangian History DIA is the correlation-time of a fluid particle along a trajectory. In the present article we will show how our theory is related to the DIA. This will in particular show why a certain number of terms which appeared in the original formulation have to be taken equal to zero for consistency with DIA. We will also apply the approach to the problem of the diffusion of a passive scalar. In the framework of the EDQNM theory, we will propose a model for the scalar flux spectrum, which significantly simplifies the level of complexity with respect to the models proposed in previous works \cite{Herr,Ulitsky,Bos2005,Bos2007-1}. 
%Results will be compared to results from classical EDQNM theory and the assumptions contained in the latter will be discussed in the light of the 
%new closure.

\section{Lagrangian dynamics, Markovianization and determination of the correlation time-scale\label{sec:Lagr}}

\begin{figure}
\begin{center}
\setlength{\unitlength}{.5\textwidth}
\includegraphics[width=1\unitlength]{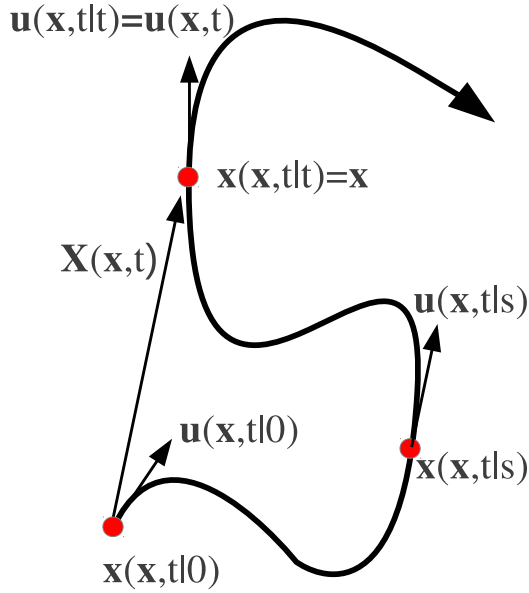}~
\caption{\label{Fig:LHDIA}  Illustration of the generalized velocity, Lagrangian position vector and displacement vector. }
\end{center}
\end{figure}

Let us introduce the spectral tensor, which is the key quantity in the statistical description of a turbulent flow.
\begin{eqnarray}\label{PhiEul}
\Phi_{ij}(\bm k,t,s)&=&\mathcal{F}_{(\bm r)}\left\{\overline{u_i(\bm x,t)u_j(\bm x+\bm r,s)}\right\}\\
&=&\overline{u_i(\bm k,t)u_j^*(\bm k,s)}
\end{eqnarray}
with $\mathcal{F}_{(\bm r)}$ a Fourier transform with respect to $\bm r$, $u_i(\bm k,t)$ the Fourier transform of the velocity and the overline denoting an ensemble average. The conjugate $u_j^*(\bm k,s)=u_j(-\bm k,s)$ since the velocity is a real quantity in physical space.  In incompressible, mirror-symmetric, isotropic turbulence this quantity is entirely determined by a scalar function, $E(k,t,s)$, the energy spectrum, by the relation
\begin{equation}
\Phi_{ij}(\bm k,t,s)=\frac{E(k,t,s)}{4\pi k^2}P_{ij}(\bm k),
\end{equation}
with
\begin{equation}
P_{ij}(\bm k)=\delta_{ij}-\frac{k_ik_j}{k^2}.
\end{equation}
The goal of the present work is to obtain a statistical model for the dynamics of $E(k,t,t)\equiv E(k,t)$. The original DIA theory was based on the above quantities, or representatives, yielding a $k^{-3/2}$ inertial range scaling. This behavior was related to the fact that an Eulerian description does not distinguish between the decorrelation of the velocity modes by the effects of pressure and viscosity on one hand and the sweeping of the small scales by the large scales on the other hand. Sweeping does not contribute to the energy transfer among scales which is the physical mechanism behind the $k^{-5/3}$ prediction. Indeed an Eulerian observer would have no way to see if the decorrelation is due to sweeping or not. To correctly capture the dynamics one needs a Lagrangian description. In other words, one needs to change the representative from (\ref{PhiEul}) to a Lagrangian formulation. 
%NEW============================
Kraichnan reformulated his theory in Lagrangian coordinates \cite{Kraichnan65}. Hereto he introduced the generalized velocity, 
 $u_i(\bm x,t|s)$, defined as the velocity measured at time $s$ on the trajectory which passes at $t$ through $\bm x$. The first time argument is called the labeling time, and the second, the measuring time.  This velocity is illustrated in Figure \ref{Fig:LHDIA}. 
We can associate the Lagrangian position function to this velocity, $x_i(\bm x,t|s)$, defined as the position of the fluid particle at time $s$ on the trajectory which passes through $\bm x$ at time $t$. These quantities are related to the Eulerian velocity and coordinate system by,
\begin{eqnarray}
\bm u(\bm x,t|t)&=&\bm u(\bm x,t),\\
\bm x(\bm x,t|t)&=&\bm x.
\end{eqnarray}
The key quantity that we will use in the following is the Eulerian displacement field $\bm X$ related to the generalized velocity field through the relation 
\begin{equation}\label{eq:Xintu}
X_i(\bm x,t)=\int_0^t u_i(\bm x,t|s)ds=x_i(\bm x,t|t)-x_i(\bm x,t|0).
\end{equation}
The displacement field is an Eulerian field, which measures at each point $\bm x$ in a fixed, Eulerian, coordinate system the displacement with respect to its original position (at $t=0$) of the particle which is at point $\bm x$ at time $t$ (see for example the discussion in reference \cite{Corrsin1962}). The displacement vector is also illustrated in Figure \ref{Fig:LHDIA}. Its evolution is given by
\begin{equation}
\frac{\partial X_i(\bm x,t)}{\partial t} +u_j(\bm x,t)\frac{\partial}{\partial x_j}X_i(\bm
x,t)=u_i(\bm x,t).
\end{equation}
Note that since the displacement field is a quantity measured in an Eulerian coordinate system, the evolution equation contains a convective derivative. 

Using the generalized velocity, the spectral velocity-tensor now contains four time-arguments,
\begin{equation}
\Phi_{ij}(\bm k,t|s;t'|s')=
\overline{u_i(\bm k,t|s)u_j^*(\bm k,t'|s')}.
\end{equation}
However, in the Lagrangian Renormalized Approximation \cite{Kaneda81} or in the Abridged Lagrangian History DIA \cite{Kraichnan65}, the resulting model equations only contain quantities of the form
\begin{equation}\label{eq:Phittts}
\Phi_{ij}(\bm k,t|t;t|s)=
\overline{u_i(\bm k,t|t)u_j^*(\bm k,t|s)},
\end{equation}
\ie, one of the velocities is evaluated at a fixed point, since the labeling and measuring time coincide, and the other is moving along its Lagrangian trajectory. The resulting model thus only contains two-time quantities, like the Eulerian DIA, but the time-difference is evaluated along a particle trajectory. The difference between LRA and the Abridged Lagrangian History DIA is that in the former, time derivatives are taken with respect to the measuring time, whereas in the the latter these derivatives are taken with respect to the labeling time.

We note that the Lagrangian velocity is not solenoidal. The Eulerian velocity is incompressible so that $\Phi_{ij}(\bm k,t|t;t|s)$ is solenoidal in its first index $i$. Isotropy and mirror-symmetry imply then that the tensor $\Phi_{ij}(\bm k,t|t;t|s)$ is proportional to $P_{ij}(\bm k)$. We can therefore define the spectrum,
\begin{eqnarray}
\Phi_{ij}(\bm k,t|t;t|s)=\frac{E(k,t|t;t|s)}{4\pi k^2}P_{ij}(\bm k),
\end{eqnarray}
but this relation is not valid for $\Phi_{ij}(\bm k,t|s;t'|s')$. Since only two-time quantities are used in the following, the spectral tensors and spectra will not be denoted by the four time-arguments, but will be denoted by their short-hand notation,
\begin{eqnarray}
\Phi_{ij}(\bm k,t|s)&\equiv&\Phi_{ij}(\bm k,t|t;t|s)\\
E(k,t|s)&\equiv& E(k,t|t;t|s).
\end{eqnarray}
This notation should not introduce confusion, because we will only consider spectra in which three of the time-arguments are evaluated at $t$.  Note that the Lagrangian and Eulerian representative coincide when $s=t$,
\begin{eqnarray}
E(k,t|t)=E(k,t,t)=E(k,t).
\end{eqnarray}
The Abridged Lagrangian History DIA constitutes a closed set of equations, yielding a $k^{-5/3}$ inertial range behavior.  The equation governing the evolution of $E(k,t)$ is 
\begin{eqnarray}\label{eq:Lin}
\left[\frac{\partial }{\partial t}+2\nu k^2\right]E(k,t)=\iint_{\Delta}\int_0^t[xy+z^3]pE(q,t|s)\left[k^2 G(k,t|s)E(p,t|s)\right.\nonumber\\
\left. -p^2G(p,t|s)E(k,t|s)\right]ds\frac{dp dq}{pq},
\end{eqnarray}
in which $G(k,t|s)$ is the response function measured along a trajectory. The symbol $\Delta$ indicates the domain in the $pq$-plane in which
$k,p,q$ can form a triangle (in other words $|p-q|\leq k \leq|p+q|$) and $x,y,z$ are given by
\begin{eqnarray}
x=-p_iq_i/(pq)\nonumber\\
y=k_iq_i/(kq)\nonumber\\
z=k_ip_i/(kp).\label{eq:xyz}
\end{eqnarray}
The LRA equations differ from (\ref{eq:Lin}) in that the RHS contains quantities of the form $G(k,s|t)$ and $E(k,s|t)$ instead of quantities of the form $G(k,t|s)$ and $E(k,t|s)$. In other words, the labeling and measuring times are inverted in the two-time quantities. We will not consider the LRA in the following. In the Abridged Lagrangian History DIA, equation (\ref{eq:Lin}) is part of a system of three equations, the other two giving the evolution for  $E(k,t|s)$ and $G(k,t|s)$, respectively. The equations, although simpler than the original Lagrangian History DIA equations, remain rather complicated and depend on the entire time-history of the flow. Our goal is to simplify the description to obtain a model that does not depend on the entire time-history of the flow, while retaining as much as possible from the original description. We will now explain how we obtain such a model.

We introduce the correlation function $R(k,t|s)$ as,
\begin{eqnarray}\label{eq:Rkts}
R(k,t|s)=\frac{E(k,t|s)}{E(k,t,t)},
\end{eqnarray}
which allows to rewrite (\ref{eq:Lin}) as
\begin{eqnarray}\label{eq:Lin2}
\left[\frac{\partial }{\partial t}+2\nu k^2\right]E(k,t)=\iint_{\Delta}[xy+z^3]pE(q,t)\left[\hat\Theta(kpq)k^2 E(p,t)\right.\nonumber\\
\left.-\hat\Theta(pkq)p^2G(p,t)E(k,t)\right]\frac{dp dq}{pq},
\end{eqnarray} 
with 
\begin{eqnarray}\label{Theta12}
\hat\Theta(kpq)=\int_0^t G(k,t|s)R(p,t|s)R(q,t|s)ds
\end{eqnarray} 
At this point we have not introduced any assumptions with respect to the Abridged Lagrangian DIA. A first assumption that we make is 
\begin{eqnarray}\label{eq:FDT}
 R(k,t|s)=G(k,t|s) \textrm{~~for~~} t>s, 
 \end{eqnarray}
which is a fluctuation-dissipation hypothesis, which leads to
\begin{eqnarray}\label{eq:Lin3}
\left[\frac{\partial }{\partial t}+2\nu k^2\right]E(k)=\iint_{\Delta}\Theta(kpq)[xy+z^3]pE(q)\left[k^2  E(p)-p^2E(k)\right]\frac{dp dq}{pq},
\end{eqnarray}
with
\begin{eqnarray}\label{eq:Theta3}
\Theta(kpq)=\int_0^t G(k,t|s)G(p,t|s)G(q,t|s)ds.
\end{eqnarray}
The fluctuation-dissipation theorem can be proven to hold rigorously in equilibrium thermodynamics, but in turbulence, corresponding to an out-of-equilibrium dissipative system, this is not the case \cite{Kraichnan2000}. It is observed in \cite{Kraichnan66} that in the Abridged Lagrangian History DIA, the fluctuation-dissipation relation is violated. In the Lagrangian Renormalized Approximation, however, which is different from 
Lagrangian History DIA with respect to the 
choice of the representative, the fluctuation-dissipation relation holds surprisingly \cite{Kaneda81,Kaneda1986}. The precise form of the relation between $G$ and $R$ is therefore not uniquely determined and can have important consequences for the Markovian models which result from the assumption, in particular when wave-like phenomena are considered \cite{Bowman1993}.

In the above description, only the {\it triad time-scale} $\Theta(kpq)$ depends on the time-history of the flow, through the temporal dependence of the response function $G(k,t|s)$. Therefore, by modeling the time-dependence of $G(k,t|s)$, or equivalently $R(k,t|s)$, a single-time description can be obtained. It is at this point that we will proceed in simplifying the description, by introducing specific assumptions on the time-dependence of $R(k,t|s)$.
A plausible assumption is that two-time correlations decay exponentially with the time-difference. 
%(for numerical support, see for example \cite{Gotoh2} (check if this is the correct reference!!!)). 
This cannot be correct for very small times since the time-dependence should be differentiable at zero time difference and such a discontinuity can influence the results in the case of compressible turbulence \cite{Fauchet1999}.  A Gaussian time-dependence could be assumed to get correct behavior around zero. It was however shown that in the incompressible case the results are relatively insensitive to the exact choice \cite{Kraichnan1971}, since only the integrals over the exponential are required and the exact form of the time-correlation does not appear in the final expressions.  For convenience we will assume therefore an exponential time-dependence here:
\begin{equation}\label{PhiMarkov}
R(k,t|s)=\exp\left[-\frac{|t-s|}{\tau(k,t)}\right],
\end{equation}
%=======================
Note that also the choice that $\tau$ is here a function of $t$ (and not of $s$) is a nontrivial assumption. We can integrate both sides of this expression with respect to $s$ to obtain an expression for $\tau(k,t)$,
\begin{equation}
\tau(k,t)=\int_{t_0}^{t}R(k,t|s)ds\left(1- \exp\left[-\frac{|t-t_0|}{\tau(k,t)}\right]\right)^{-1}.
\end{equation}
For the sake of simplicity, we will consider here the case in which the turbulence has been created a time long before we evaluate it, by taking the limit of $t_0$ tending towards  $-\infty$. This gives the expression
\begin{equation}\label{eq:tau5}
\tau(k,t)=\int_{-\infty}^{t}R(k,t|s)ds=\int_{-\infty}^{t}\frac{E(k,t|s)}{E(k,t|t)}ds.
\end{equation}
The essential step to get a closed model expression for $\tau(k,t)$, which was proposed in \cite{Bos2006}, was the explicit integration of $E(k,t|s)$ with respect to $s$. 
%===============
% It is this step which leads to a formulation closed in Eulerian quantities only. 
Using relation (\ref{eq:Xintu}), we can evaluate the integral (\ref{eq:tau5}),
\begin{eqnarray}
\int_{-\infty}^{t}E(k,t|s)ds&=&2\pi k^2 \mathcal{F}_{(\bm r)}\left\{\int_{-\infty}^{t}\overline{u_i(\bm x,t|t)u_i(\bm x+\bm r,t|s)}ds\right\} \nonumber\\
&=&2\pi k^2 \mathcal{F}_{(\bm r)}\left\{\overline{u_i(\bm x,t|t)X_i(\bm x+\bm r,t)}\right\}\nonumber\\
&=&F(k,t,t),
\end{eqnarray}
in which we introduced a new quantity, the velocity-displacement cross-correlation spectrum $F(k,t,s)$, 
\begin{eqnarray}\label{SpecTensF}
\overline{u_i(\bm k,t) X_j(-\bm k,s)}=\frac{F(k,t,s)}{4\pi k^2}P_{ij}(\bm k),
\end{eqnarray}
and we will write
\begin{eqnarray}
F(k,t,t)= F(k,t).
\end{eqnarray}
At this point we have obtained an Eulerian single-time expression for the Lagrangian time-scale,
\begin{equation}\label{taubdqnm}
\tau(k,t)=\frac{F(k,t)}{E(k,t)}.
\end{equation}
If the equation for $E(k,t)$ is completed by an equation for $F(k,t)$, we obtain a closed set of equations, if the equation for $F(k,t)$ does not introduce other unknown quantities. In \cite{Bos2006} this was accomplished within the framework of the EDQNM closure. In the next section we will obtain similar expressions, using the DIA formalism which will clarify uncertainties in the original derivation. It was observed by Kraichnan \cite{Kraichnan65} that the evolution of $E(k,t|t)$ in the Abridged Lagrangian History DIA is given by the same equation as $E(k,t)$ in the Eulerian DIA. The only difference is that the two-time quantities are now evaluated along particle trajectories. We will use this as an {\it ansatz} in the following and we will derive the Eulerian DIA expression for $F(k,t)$, in which we will evaluate the two-time correlations using Kraichnan's generalized Lagrangian description. In section \ref{sec:Num} the resulting equations will be numerically integrated and results will be presented for 
both the dynamics of the kinetic energy spectrum and the dynamics of a passive scalar advected by turbulent flow.
%===================

\section{Derivation of the evolution equation for the velocity-displacement cross-correlation}\label{Derivation}

We will focus in this section on the derivation of the evolution equation for the wavenumber spectrum of the correlation between the velocity $u_i$ and the displacement vector $X_i$ of a fluid particle. The DIA formalism will be used to evaluate the triple correlations which appear in this equation. A Markovianization procedure is applied to the resulting expressions to obtain a single-time expression.  We will try to give a self-contained description of all the operations used in this derivation. However, for a justification of the assumptions in the DIA we refer to the original work \cite{KraichnanDIA} or to later works that discuss the justification of the Lagrangian variants of the DIA \cite{Kraichnan1977,Kaneda81,Kida1997,Kaneda2007}.
%, and to the textbook by \cite{Leslie}, which is entirely dedicated to DIA and its derivatives. 
The statistically isotropic and mirror-symmetric case will be considered but in principal this derivation could also be performed for the anisotropic case. 

The starting point are the equations for $u_i$ and $X_i$,
\begin{eqnarray}
\frac{\partial  u_i(\bm x,t)}{\partial t}+u_j(\bm x,t)\frac{\partial}{\partial x_j}u_i(\bm
x,t)=-\frac{\partial p(\bm x,t)}{\partial x_i}+\nu \frac{\partial^2}{ \partial x_j^2}u_i(\bm x,t)\nonumber\\
\frac{dX_i(\bm x,t)}{dt}\equiv\frac{\partial X_i(\bm x,t)}{\partial t} +u_j(\bm x,t)\frac{\partial}{\partial x_j}X_i(\bm
x,t)=u_i(\bm x,t),          \label{eq:XFIRST}
\end{eqnarray}
which in Fourier space are written as
\begin{eqnarray}\label{eqNSX}
[\frac{\partial}{\partial t}+\nu k^2]u_i(\bm
k,t)=-\frac{i}{2}P_{ijm}(\bm k)\iint \delta_{\bm k-\bm p-\bm q}
u_j(\bm p,t)u_m(\bm q,t)d\bm p d\bm q\nonumber\\
\frac{\partial X_i(-\bm k,t) }{\partial t} =i\bm k_j \iint \delta_{\bm k-\bm p-\bm q}u_j(-\bm p,t)X_i(-\bm q,t)d\bm p d\bm q+u_i(-\bm k,t),
\end{eqnarray}
with
\begin{eqnarray}
P_{ijm}(\bm k)=k_jP_{im}(\bm k)+k_mP_{ij}(\bm k). 
\end{eqnarray}
Multiplying the first equation by $X_i(-\bm k,t)$, the second by
$u_i(\bm k,t)$, summing and ensemble averaging gives 
\begin{eqnarray}\label{eqF01}
[\frac{\partial}{\partial t}+\nu k^2]F(k,t,t)=T^u(k,t)+T^X(k,t)+E(k,t,t)
\end{eqnarray}
with
\begin{eqnarray}\label{eqTuip}
T^u(k,t)=-i\pi k^2 P_{ijm}(\bm k)\iint \delta_{\bm k-\bm p-\bm q}\overline{u_j(\bm p,t)u_m(\bm q,t)X_i(-\bm k,t)}d\bm p d\bm q\nonumber\\
T^X(k,t)=2i\pi k^2\bm k_j \iint \delta_{\bm k-\bm p-\bm q}\overline{u_i(\bm k,t) u_j(-\bm p,t)X_i(-\bm q,t)}d\bm p d\bm q.
\end{eqnarray}
The DIA is a perturbation method in which the small parameter is the influence of a single triad interaction on the velocity field, compared to the velocity field obtained by the interaction of all other triad interactions. In practice, DIA expressions are obtained by replacing, in the correlations that one is interested in, the velocity field by 
\begin{eqnarray}\label{expansion}
%u_i\rightarrow u_i^{(0)}+u_i^{(1)}+ u_i^{(2)}+...
u_i\rightarrow u_i^{(0)}+u_i^{(1)}
\end{eqnarray}
and using a similar expression for the displacement vector $X_i$. The part $u_i^{(0)}$ is the velocity field that would exist if all triad interactions are retained, except for one that we will call $(\bm k, \bm p, \bm q)$. Since in a homogeneous flow an infinite number of triads interact, $u_i^{(0)}$ is supposed to be close to the actual velocity field. The contribution $u_i^{(1)}$ corresponds to the velocity field induced by the interaction $(\bm k, \bm p, \bm q)$ only. The justification of this approach can be interpreted in terms of the coupling strength between the different modes that should be sufficiently small. If this coupling is weak enough and the number of modes tends to infinity, the DIA procedure should give good results. This was tested in reference \cite{Goto2002} on a toy-model in which both the coupling strength and the number of modes could be varied.
%The second order $u_i^{(2)}$ is similarly the contribution to $u_i^{(1)}$ by a single triad $(\bm k', \bm p', \bm q')$. 
%In the following only the first order perturbation is needed to obtain nontrivial results. 

After replacing all $u$'s and $X$'s in $T^u$ by
  (\ref{expansion}) and developing, we find that
\begin{eqnarray}\label{TuExp}
T^u(\bm k)=-i\pi k^2 P_{ijm}(\bm k)\iint \delta_{\bm k-\bm p-\bm q}
\underbrace{\left[\overline{u_j^{(0)}(\bm p)u_m^{(0)}(\bm q)X_i^{(0)}(-\bm k)}\right.}_{\textrm{zeroth order}}
+
\nonumber\\
\underbrace{\overline{u_j^{(1)}(\bm p)u_m^{(0)}(\bm q)X_i^{(0)}(-\bm k)}
+\overline{u_j^{(0)}(\bm p)u_m^{(1)}(\bm q)X_i^{(0)}(-\bm k)}
+\overline{u_j^{(0)}(\bm p)u_m^{(0)}(\bm q)X_i^{(1)}(-\bm k)}}_{\textrm{first order}}\nonumber\\
\left. +\textrm{~second and higher order} \right]d\bm p d\bm q,~~~~~~~~~~~~~~~~~
\end{eqnarray} 
in which time-arguments were omitted to shorten the expression. A
similar expression is obtained for $T^X$. In the DIA only the first non-vanishing order is retained in addition to the zeroth order contribution. Higher order contributions should contribute insignificantly if the contribution of a single triad is indeed infinitesimally small. For the triple correlations considered here, the zeroth order contribution (first term on the RHS) in the above expression is equal to zero. The first nonvanishing contribution is the first order contribution. 
%Now the velocities $u^{(1)}$ are obtained by retaining the nonlinear
%  response of an infinitesimal disturbance of the Navier-Stokes
%  equation. 
The Direct Interaction contributions can be written for, for example $u_j(\bm p,t)$ and $X_i(\bm k,t)$, as
\begin{eqnarray}\label{up'}
u_j^{(1)}(\bm p,t)&=&-i P_{jab}(\bm p) \int_0^t G(\bm p,t,s)\left[u_a^{(0)}(\bm k,s)u_b^{(0)}(-\bm q,s)\right]ds  \nonumber\\
X_i^{(1)}(-\bm k,t)&=&ik_a\int_0^t G^X(\bm k,t,s)\left[u_a^{(0)}(-\bm p,s)X_i^{(0)}(-\bm q,s)
+u_a^{(0)}(-\bm q,s)X_i^{(0)}(-\bm p,s)\right]ds, \nonumber\\
\end{eqnarray} 
%Note that these responses always verify triad interactions
%$\bm k=\bm p+\bm q$. A perturbed $\bm k$ interacts with $\bm p$ and
%$\bm q$ and equally a perturbed $\bm p$ interacts with $-\bm q$ and
%$\bm k$ and so forth. 
in which $G(\bm k,t,s)$ and  $G^X(\bm k,t,s)$ are the response function of the velocity and displacement field, respectively. These are unknown quantities at this point.

These direct interaction contributions are substituted into expression (\ref{TuExp}). To illustrate the derivation of the final expression we take the first term on the second line of expression (\ref{TuExp}), which we will denote $T_1(k)$, and we will work out the final expression for this term. Upon substituting (\ref{up'}) in this term we obtain,
\begin{eqnarray}\label{T1k}
T_1(k)=- \pi k^2\int_{0}^{t}\iint \delta_{\bm k-\bm p-\bm q}G(\bm p,t,s)P_{ijm}(\bm k)P_{jab}(\bm p)\times\nonumber\\
\overline{ u_a^{(0)}(\bm
k,s)u_b^{(0)}(-\bm q,s) u_m^{(0)}(\bm q,t)X_i^{(0)}(-\bm k,t)} d\bm p d\bm q ds.
\end{eqnarray}
This expression is exactly what would be obtained using the quasi-normal approach, for a particular choice of $G$, the viscous Green-function. However, DIA does not only take into account viscous decorrelation, but also non-linear decorrelation and the DIA expression for $G$ will therefore differ from the viscous Green-function. 
In the limit, which we will consider, in which the flow is constituted of a very large number of modes,  we can write the correlation in (\ref{T1k}) as a product of double correlations (note that this is no assumption of Gaussianity of the velocity modes, as for example discussed in \cite{Goto2002}),
\begin{eqnarray}\label{T1k-3}
\overline{ u_a^{(0)}(\bm k,s)u_b^{(0)}(-\bm q, s) u_m^{(0)}(\bm q,t)X_i^{(0)}(-\bm k,t)}=P_{bm}(\bm q)\frac{E(q,t,s)}{4\pi q^2}P_{ai}(\bm k)\frac{F(k,s,t)}{4\pi k^2},\nonumber\\
\end{eqnarray}
in which we used the isotropic expressions for the velocity and velocity-displacement tensors.
%Once we have substituted these expressions, we invoke the weak dependence
%hypothesis \cite{KraichnanDIA} which basically corresponds to a Quasi-Normal approximation of the two-time quadruple correlations. Note that the difference with the basic  Quasi-Normal approximation is that the assumption is here applied to the two-time velocity correlations which appear as a result of the perturbation method. In this expression, we therefore replace
%\begin{eqnarray}\label{QN}
%\overline{ u_a^{(0)}(\bm p',s)u_b^{(0)}(\bm p'', s) u_m^{(0)}(\bm q,t)X_i^{(0)}(-\bm k,t)}\rightarrow\Phi_{ab}(\bm p',s,s)\delta_{\bm p'+\bm p''}\Phi_{mi}(\bm q,t,t)\delta_{\bm q-\bm k}+\nonumber\\
%\Phi_{am}(\bm q,s,t)\delta_{\bm p'+\bm q}\Phi^X_{bi}(\bm k,s,t)\delta_{\bm p''-\bm k}+
%\Phi^X_{ai}(\bm k,s,t)\delta_{\bm p'-\bm k}\Phi_{bm}(\bm q,s,t)\delta_{\bm p''+\bm q}
%\end{eqnarray}
%in which we used that in statistically homogeneous flows
%\begin{eqnarray}
%\overline{u_i(\bm k, t) u_j(\bm p,s)}=\Phi_{ij}(\bm k,t,s)\delta_{\bm k+\bm p}
%\end{eqnarray}
Introducing this in the above expression we find
%we can simplify the expressions by integrating over $\bm p'$ and $\bm p''$, which shows that the first contribution in (\ref{QN}) vanishes, giving,
\begin{eqnarray}
T_1(k)=- \pi k^2\int_{0}^{t}\iint \delta_{\bm k-\bm p-\bm q}
G(\bm p,t,s)P_{ijm}(\bm k)P_{jab}(\bm p)
P_{bm}(\bm q)P_{ai}(\bm k)\frac{E(q,t,s)}{4\pi q^2}\frac{F(k,s,t)}{4\pi k^2}
d\bm p d\bm q ds,\nonumber\\ \label{usymm}
\end{eqnarray}
%Using isotropic forms for the spectral tensor and the velocity-displacement correlation tensor,
%\begin{eqnarray}\label{SpecTens2}
%\Phi_{ij}(\bm k,t,s)=P_{ij}(\bm k)\frac{E(k,t,s)}{4\pi k^2},\qquad\qquad
%\Phi^X_{ij}(\bm k,t,s)=P_{ij}(\bm k)\frac{F(k,t,s)}{4\pi k^2}
%\end{eqnarray}
which can be written as
\begin{eqnarray}\label{eqTabc3}
T_1(k)=- \pi k^2\iint \delta_{\bm k-\bm p-\bm q}\int_0^t ds~ G(p,t,s)kp~f_1(k,p,q)E(q,t,s)F(k,s,t)(16\pi^2
k^2 q^2)^{-1}d\bm p d\bm q\nonumber\\
\end{eqnarray}
with $f_1(kpq)=(kp)^{-1}P_{ajm}(\bm k)P_{jab}(\bm p)P_{bm}(\bm q)$ in which we used that $P_{ijm}(\bm k)P_{ai}(\bm k)=P_{ajm}(\bm
k)$. The convolution integral can be rewritten as a function of the
wavenumber only. The details of this simplification are given in the
appendix of the book by Leslie \cite{Leslie}. The resulting expression
is 
\begin{eqnarray}
\iint \delta_{\bm k-\bm p-\bm q}d\bm p d\bm q\rightarrow\iint_\Delta 2\pi pqk^{-1}dp dq
\end{eqnarray}
in which $\Delta$ indicates the domain in the $pq$-plane in which
$k,p,q$ can form a triangle (in other words $|p-q|\leq k \leq|p+q|$). This
yields 
\begin{eqnarray}\label{eqTabc4}
T_1(k)&=&- \frac{1}{8}\int_{\Delta}\int_0^t ds~ G(p,t,s)f_1(k,p,q)p^3E(q,t,s)F(k,s,t)
 \frac{dp}{p}\frac{dq}{q}.
\end{eqnarray}
We see that the unknown quantities in this description are the
$2$-time quantities $E(k,t,s),F(k,s,t),G(k,t,s)$. It is here that we replace these quantities by their analogues defined on Lagrangian trajectories,
\begin{eqnarray}
E(k,t,s)\rightarrow E(k,t|s)\nonumber\\
G(k,t,s)\rightarrow G(k,t|s)\nonumber\\
F(k,t,s)\rightarrow F'(k,t|s)\nonumber\\
F(k,s,t)\rightarrow F''(k,t|s),
\end{eqnarray}
with
\begin{eqnarray}
F'(k,t|s)=2\pi k^2\overline{u_i(\bm k,t) X_i(-\bm k,t|s)}\nonumber\\
F''(k,t|s)=2\pi k^2\overline{u_i(\bm k,t|s) X_i(-\bm k,t)} 
\end{eqnarray}
in which the difference between $F'$ and $F''$ arises due to the asymmetric character of the velocity-scalar displacement correlation.
%%%%%%%
%The DIA proposes a
%method to write an equation for the response function, which allows to
%close the set of equations for these two-time quantities. 
%This yields a description which involves the time-history of the flow. Such a description needs large numerical resources (at infinite times, if the time-history is not truncated, an infinite time-history should be
%taken into account). 
%In order to simplify the description to a single-time description, we need an assumption about the time-dependence of $E(k,t,s),F(k,t,s),G(k,t,s)$. 
As in section \ref{sec:Lagr}, expression (\ref{PhiMarkov}), we will use the assumption that the two-time quantities decay exponentially in time by posing
\begin{eqnarray}
G(k,t|s)=H(t-s)\exp[-(t-s)\eta(k)]\\
G^X(k,t|s)=H(t-s)\exp[-(t-s)\eta^X(k)]\\
E(k,t|s)=E(k,t)[G(k,t|s)+G(k,s|t)]\label{responseE}\\
F'(k,t|s) =F(k,t)[G(k,s|t)+G(k,t|s)]\label{responseF}\\
F''(k,t|s)=F(k,t)[G^X(k,s|t)+G^X(k,t|s)]\label{responseF2}.
\end{eqnarray}
The Heaviside functions $H(t)$ appear since the response functions correspond to the response at time $s$ to a perturbation applied at time $t$. There is obviously no response before the perturbation is applied. %Note that expression (\ref{responseF}) depends on the order of the time-arguments, since in one case the velocity is evaluated along a trajectory and in the other case the displacement vector. 
Note further that (\ref{responseE}), (\ref{responseF}) and (\ref{responseF2}), which link the response functions to the velocity time-correlations, are, like expression (\ref{eq:FDT}), a specific form of the fluctuation-dissipation theorem.  Using the above expressions, the time-integral will, in the long-time limit, yield 
\begin{eqnarray}\label{eqTabc5}
\int_{-\infty}^t ds~ G(p,t|s)E(q,t|s)F''(k,t|s)=\Theta^F(kpq)E(q,t)F(k,t)\nonumber\\
\end{eqnarray}
with 
\begin{eqnarray}
\Theta^F(kpq)=\frac{1}{\eta^X(k)+\eta(p)+\eta(q)}
\end{eqnarray}
We repeat the foregoing analysis for the remaining two terms in $T^u_{ip}$
and the three terms of $T^X_{ip}$. The Direct Interaction contribution stemming from the $X_i$
in each term yields two contributions, whereas every $u_i$ yields only one contribution (\emph{c.f.} the symmetry used in expression (\ref{usymm})). The total number of terms is
then eight. In the following all quantities are evaluated at time $t$ so that
we will omit the time arguments. The resulting equation is
\begin{eqnarray}\label{eqF10}
[\frac{\partial}{\partial t}+\nu k^2]F(k)=\sum_{i=1}^8 T_i(k)+E(k)
\end{eqnarray}
with
\begin{eqnarray}
T_1(k)&=&- \frac{1}{8}\int_{\Delta}\Theta^F(kpq) f_1(k,p,q)p^3E(q)F(k)
 \frac{dp}{p}\frac{dq}{q}\\
T_2(k)&=&- \frac{1}{8}\int_{\Delta}\Theta^F(kpq) f_2(k,p,q)q^3E(p)F(k)
 \frac{dp}{p}\frac{dq}{q}\\
T_3(k)&=&+\frac{1}{8}\int_{\Delta}\Theta^F(kpq) f_3(k,p,q)k^3E(p)F(q)
 \frac{dp}{p}\frac{dq}{q}\\
T_4(k)&=& +\frac{1}{8}\int_{\Delta}\Theta^F(kpq) f_4(k,p,q)k^3E(q)F(p)
 \frac{dp}{p}\frac{dq}{q}\\
T_5(k)&=&- \frac{1}{4}\int_{\Delta}\Theta^F(qpk) f_5(k,p,q)p^3E(k)F(q)
 \frac{dp}{p}\frac{dq}{q}\\
T_6(k)&=&+ \frac{1}{4}\int_{\Delta}\Theta^F(qpk) f_6(k,p,q)k^3E(p)F(q)
 \frac{dp}{p}\frac{dq}{q}\\
T_7(k)&=&- \frac{1}{4}\int_{\Delta}\Theta^F(qpk) f_7(k,p,q)q^3E(k)F(p)
 \frac{dp}{p}\frac{dq}{q}\\
T_8(k)&=&- \frac{1}{4}\int_{\Delta}\Theta^F(qpk) f_8(k,p,q)q^3E(p)F(k)
 \frac{dp}{p}\frac{dq}{q}.
\end{eqnarray}
Symmetries allow to show that $T_1(k)=T_2(k)$ and
$T_3(k)=T_4(k)$, which reduces the number of terms to six. The geometrical functions $f_1,..,f_8$ can be expressed as a function of the cosines $x,y,z$, defined in expressions (\ref{eq:xyz}).
This yields
\begin{eqnarray}
f_1&=&(kp)^{-1}P_{123}(\bm k)P_{214}(\bm p)P_{34}(\bm q) = 2(xy+z^3)\\
f_2&=&(kq)^{-1}P_{123}(\bm k)P_{24}(\bm p)P_{314}(\bm q) =  2(xz+y^3)\\
f_3&=&(k)^{-2}k_1P_{234}(\bm k)P_{13}(\bm p)P_{24}(\bm q) =  1-xyz-2y^2z^2+y^2-z^2\\
f_4&=&(k)^{-2}k_1P_{234}(\bm k)P_{23}(\bm p)P_{14}(\bm q) =  1-xyz-2y^2z^2-y^2+z^2\\
f_5&=&(kp)^{-1}k_1P_{23}(\bm k)P_{124}(\bm p)P_{34}(\bm q) =  xy+z^3-z+zx^2\\
f_6&=&(k)^{-2}k_1P_{234}(\bm k)P_{13}(\bm p)P_{24}(\bm q) =  f_3\\
f_7&=&(kq)^{-1}k_1P_{23}(\bm k)P_{13}(\bm p)q_2 =  z(x+yz)\\
f_8&=&2(kq)^{-1}k_1 P_{12}(\bm p)q_2=  2(y+xz).
\end{eqnarray}
Note that $f_4$ and $f_2$ are not needed when symmetry is used to show that
$T_1(k)=T_2(k)$ and $T_3(k)=T_4(k)$. Since $f_6=f_3$, we eventually
only need five different factors. 

The last step in obtaining closure is the determination of $\eta(k)$ and $\eta^X(k)$. Comparison of equation (\ref{responseE}) with (\ref{PhiMarkov}) shows that within the framework of our assumptions, 
\begin{equation}
\eta(k)=\tau(k)^{-1}=E(k,t)/F(k,t).
\end{equation}
Note that the current model, as it was presented in \cite{Bos2006}
defined $\eta(k)$ as
\begin{eqnarray}\label{oldtau}
\eta(k)=E(k,t)/F(k,t)+\nu k^2. 
\end{eqnarray}
Indeed it is now clear that within the framework of the assumptions
introduced in the derivation of the curent model, the viscous
frequency should not be added. It is implicitly contained in the
expression for $F(k)$.

We have obtained a closed expression, if the unknown quantity $\eta^X(k)$ is specified. We note here that the displacement is a quantity which will not diffuse along a trajectory, neither by pressure, nor by molecular effects. This suggests the choice 
\begin{equation}\label{eq:0damp}
\eta^X(k)=0, 
\end{equation}
in analogy with Lagrangian History DIA arguments for the passive scalar.  There is however a difference between the equation of the displacement vector and the passive scalar. Indeed, the equation of the displacement vector (\ref{eq:XFIRST}) contains on the RHS the velocity. This forcing is correlated with the displacement. Some analogy exists between this case and the linearly forced Burgers equation. In the latter case the triad relaxation time is directly determined by the correlation-time of the velocity, as was shown by applying the LHDIA technique to Burgers' equation \cite{Kraichnan1968-2}. In order to test the influence of this time-scale on the results, we will also perform simulations in which 
\begin{equation}\label{eq:TINT}
\eta^X(k)=\mathcal{T}^{-1}, 
\end{equation}
in which $\mathcal{T}$ is the integral timescale of the velocity field. It will be shown that this choice does not significantly change the results.

\section{Equations for the evolution of the energy spectrum and the scalar variance spectrum}

Using the above method, equations can be derived for the evolutions of the energy spectrum and of the scalar variance spectrum. The latter is defined by
\begin{eqnarray}
E_\theta(k,t,s)=2\pi
k^2 \overline{\theta(\bm k,t)\theta(-\bm k,s)}.
\end{eqnarray}
The resulting equations are identical to the ones given by the EDQNM approach or the test-field model. The only difference is the specification of the time-scale. The equations are
\begin{eqnarray}\label{eqEK}
\left[\frac{\partial }{\partial t}+2\nu k^2\right]E(k)=\iint_{\Delta}\Theta(kpq)[xy+z^3]pE(q)\left[k^2  E(p)-p^2E(k)\right]\frac{dp dq}{pq},\\
\left[\frac{\partial }{\partial t}+2\alpha k^2\right]E_\theta(k)=\iint_{\Delta}\Theta^\theta(kpq)[1-z^2]pE_\theta(q)\left[k^2 E(p)-p^2 E(k)\right]\frac{dp dq}{pq}.\label{eqEKT}
\end{eqnarray}
The time-scales which appear in these expressions are
\begin{eqnarray}
\Theta(kpq)=\frac{1}{\eta(k)+\eta(p)+\eta(q)},\qquad
\Theta^\theta(kpq)=\frac{1}{\eta^\theta(k)+\eta'(p)+\eta^{\theta}(q)}
\end{eqnarray}
Using the arguments from the last section, the timescale $\eta(k)^{-1}$ is given by expression (\ref{taubdqnm}). The time-scale associated to the decorrelation of a passive scalar on a Lagrangian trajectory can be inferred from the equation for the scalar,
\begin{equation}
\frac{d\theta}{dt}= \alpha \Delta \theta,
\end{equation}
which suggests that the scalar decorrelation is given by 
\begin{eqnarray}\label{eq:alphakk}
\eta^{\theta}(k)=\alpha k^2,
\end{eqnarray}
since on a Lagrangian trajectory the scalar only decorrelates under the action of diffusion. We note that this choice for the scalar is inherent to the choice of the representative \cite{Kraichnan1978,Herring1979,Gotoh2000}. For example, the decorrelation time-scale for the velocity is modified if the DIA is applied to the strain-rate instead of to the velocity. However, we will use the above choice, expression (\ref{eq:alphakk}).

For comparison, it is perhaps useful here to recall the expressions for the time-scales which are used in the EDQNM model \cite{Orszag}. These are 
\begin{eqnarray}\label{timeEDQNM}
\eta(k)=\lambda\sqrt{\int_0^k s^2E(s)ds}+\nu k^2, \qquad
\eta'(k)=\lambda'\sqrt{\int_0^k s^2E(s)ds}+\nu k^2, \nonumber\\
\eta^{\theta}(k)=\lambda_\theta\sqrt{\int_0^k s^2E(s)ds}+\alpha k^2.
\end{eqnarray}
The value of $\lambda$ was determined and given the value $0.36$ by comparison with the test-field model, the latter being calibrated by comparison with the DIA in a case in which DIA was assumed to give correct results. The value for $\lambda_\theta$ was given the value $0$ in \cite{Herring}, using similar arguments as above regarding the Lagrangian decorrelation of a scalar fluctuation. In the same work $\lambda'$ was given a value different from $\lambda$. The parameter $\lambda'$ appearing in $\Theta^\theta(kpq)$ was determined by comparison with experimental results for the turbulent Prandtl number and was given the value unity. The different values for $\lambda$ and $\lambda'$ in the two different time-scales are in contradiction with the outcome of DIA. However, a certain flexibility is allowed if one drops the fluctuation-dissipation theorem. Even more freedom is obtained if one changes the representative used in the analysis. From a practical point of view one might argue that the 
appearance of adjustable constants in a model allows the user to choose them in comparison with observation rather than by using physical arguments and assumptions. 
%In the next section we will compare the present model to EDQNM, taking into account these considerations.

These expressions for the time-scales used in the EDQNM closure were only recalled here for comparison. They will not be used in the remainder of this work.

\section{EDQNM expression for the spectrum of the velocity-scalar cross-correlation}

It is at this point that we want to point out a possible simplification of the EDQNM model as proposed and derived in references \cite{Herr,Ulitsky,Bos2005,Bos2007-1}. In these works the EDQNM expression was derived for the spectrum of the velocity-scalar correlation spectrum. The resulting expressions contained a time-scale of the form (\ref{timeEDQNM}). However in the light of the present derivation it seems that the level of sophistication of that model was higher than the standard EDQNM model. Indeed, the considerations in the present work suggests a much simpler expression to model the scalar flux spectrum. Using the analogy between a passive scalar generated by a mean-scalar gradient and the displacement-vector, expression (\ref{taubdqnm}) suggests the following model for the scalar flux spectrum $F_{u\theta}(k)$,
\begin{eqnarray}
F_{u\theta}(k)&=&\Gamma E(k)/\eta_F(k)\nonumber\\
&=&\Gamma E(k)\left[\lambda_F\sqrt{\int_0^k s^2 E(s) ds} +(\nu+\alpha)k^2\right]^{-1}
\label{eq:FEDQNM}
\end{eqnarray}
with $\Gamma$ the strength of the mean scalar gradient and $\lambda_F$ a model constant. This modeling would be of the level of approximation of the EDQNM model and would avoid having to program the lengthy evolution equations which appear in the previous works.

\begin{figure}
\begin{center}
\setlength{\unitlength}{.5\textwidth}
\subfigure[]{\includegraphics[width=1\unitlength]{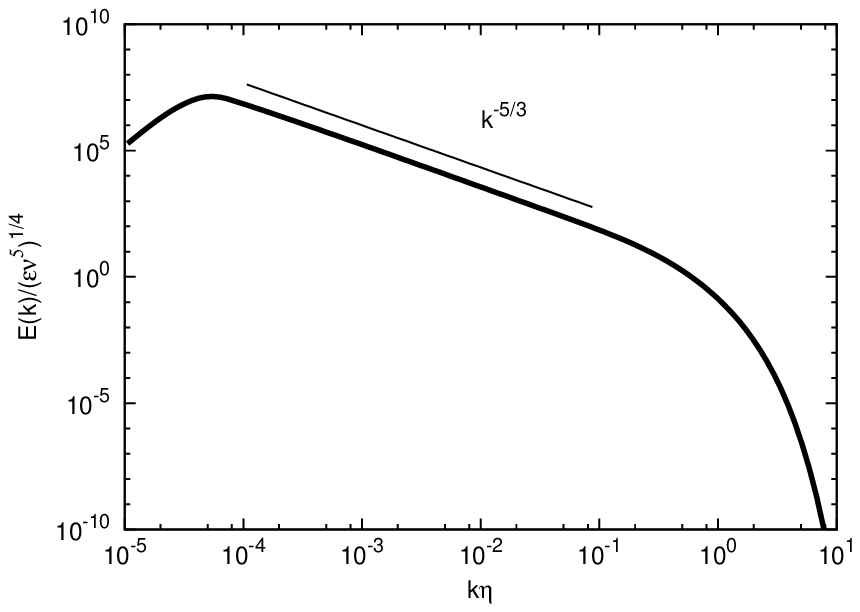}}~
\subfigure[]{\includegraphics[width=1\unitlength]{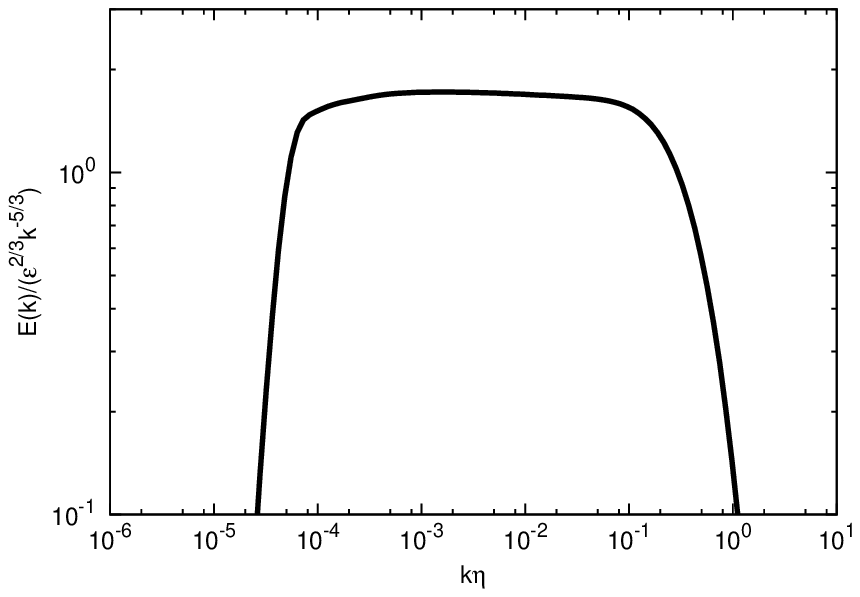}}
\caption{\label{FigEk1} Energy spectrum (a) and in compensated form (b) for decaying isotropic turbulence, modeled by the present closure approach.}
\end{center}
\end{figure}

\begin{figure}
\begin{center}
\setlength{\unitlength}{.5\textwidth}
\subfigure[]{\includegraphics[width=1\unitlength]{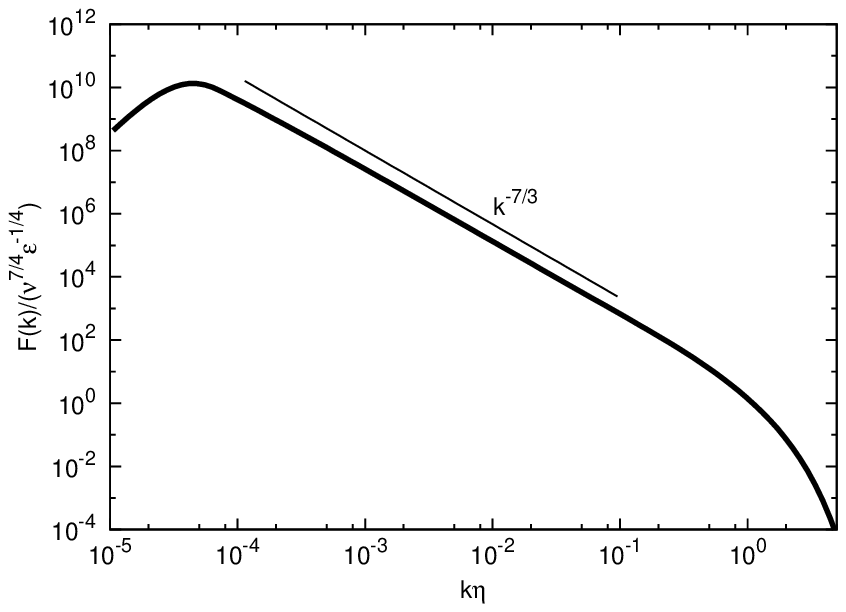}}~
\subfigure[]{\includegraphics[width=1\unitlength]{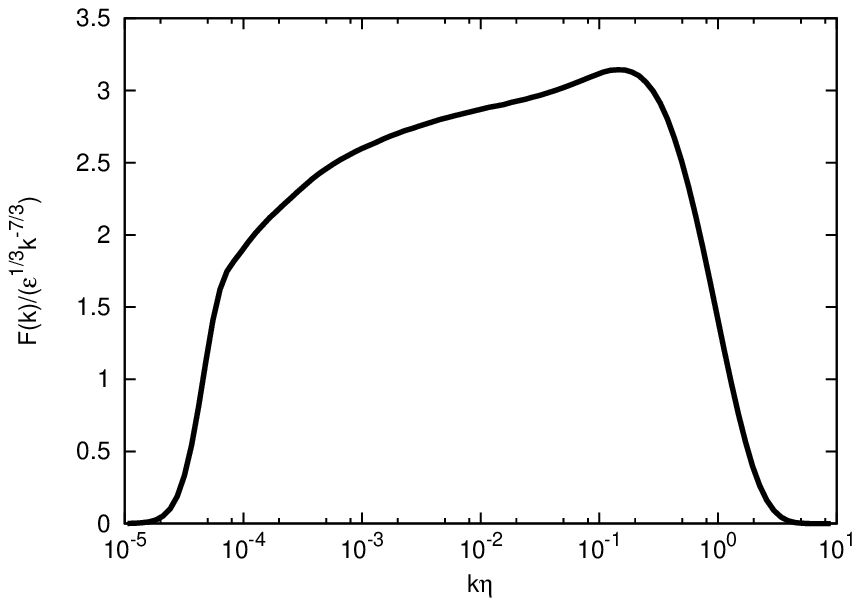}}
\caption{\label{FigEk2} Velocity-displacement cross-correlation spectrum (a) and in compensated form (b) for decaying isotropic turbulence, modeled by the present closure approach.}
\end{center}
\end{figure}

\begin{figure}
\begin{center}
\setlength{\unitlength}{.5\textwidth}
\subfigure[]{\includegraphics[width=1\unitlength]{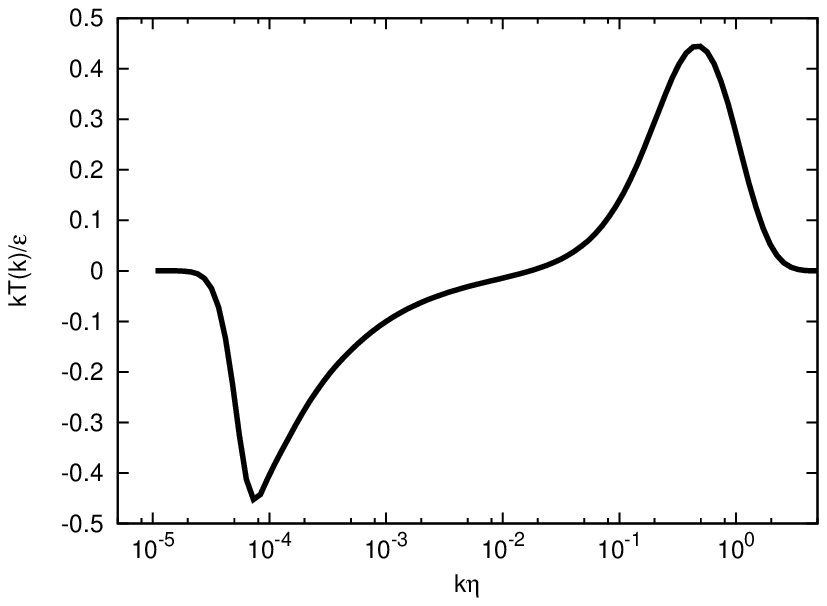}}~
\subfigure[]{\includegraphics[width=1\unitlength]{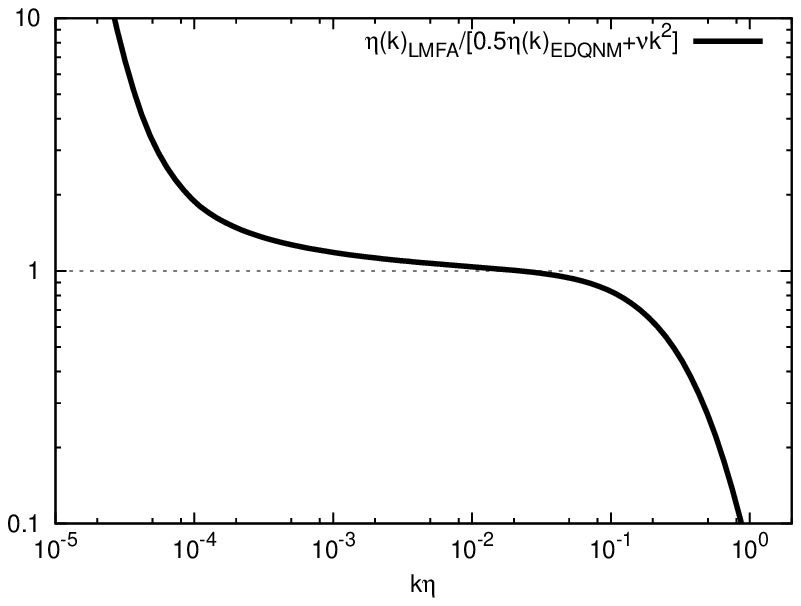}}
\caption{\label{FigEk3}  Energy transfer (a) and spectrum of the  typical (eddy-damping) time-scale compared to a Heisenberg type time-scale, usually employed in the EDQNM model (b). }
\end{center}
\end{figure}

\begin{figure}
\begin{center}
\setlength{\unitlength}{.5\textwidth}
\subfigure[]{\includegraphics[width=1\unitlength]{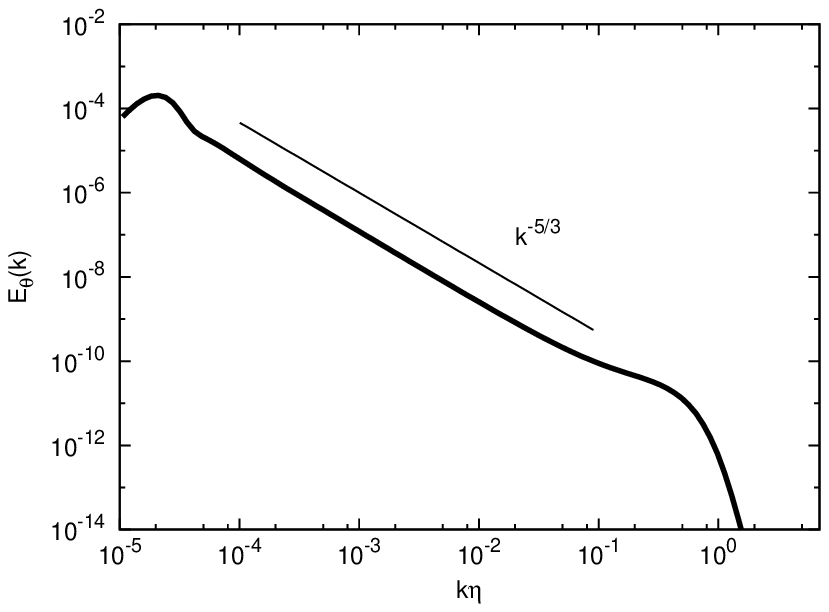}}~
\subfigure[]{\includegraphics[width=1\unitlength]{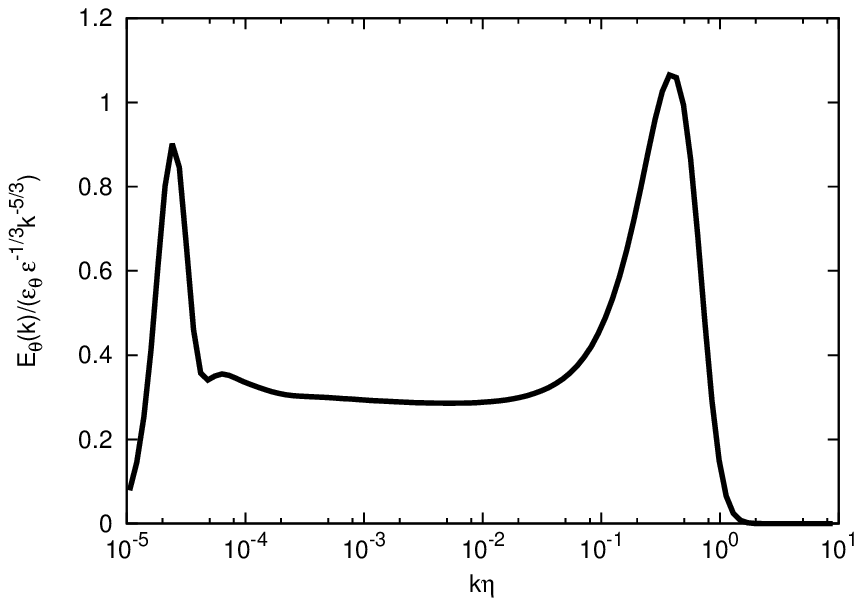}}
\caption{\label{FigETk} (a) Passive scalar variance spectrum corresponding to a decaying passive scalar at unity Schmidt number. The velocity field is characterized in Figure \ref{FigEk1}. (b) Compensated form.}
\end{center}
\end{figure}

\begin{figure}
\begin{center}
\setlength{\unitlength}{.5\textwidth}
\subfigure[]{\includegraphics[width=1\unitlength]{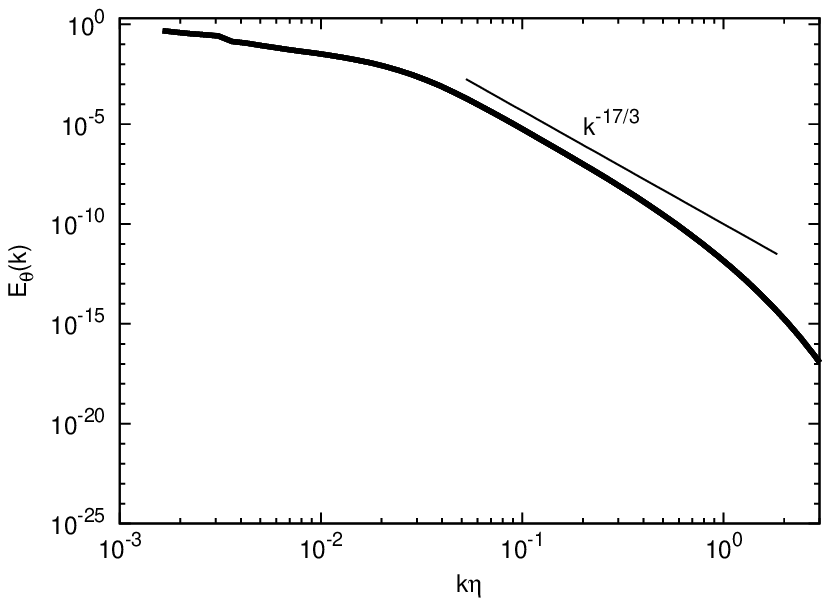}}~
\subfigure[]{\includegraphics[width=1\unitlength]{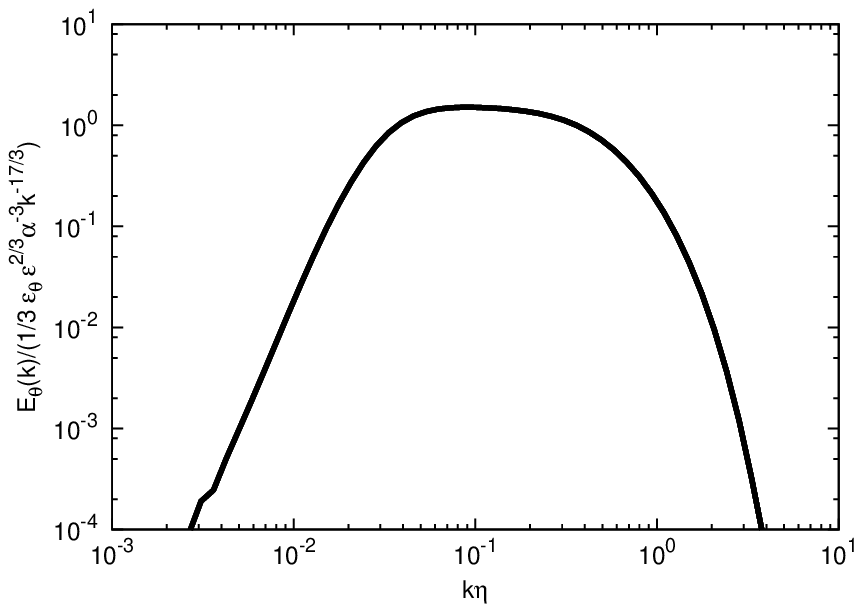}}
\subfigure[]{\includegraphics[width=1\unitlength]{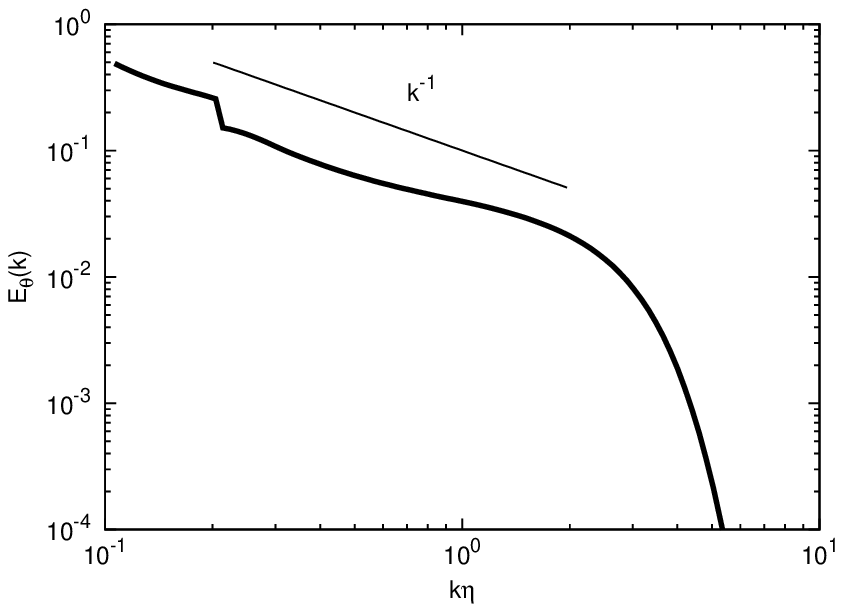}}~
\subfigure[]{\includegraphics[width=1\unitlength]{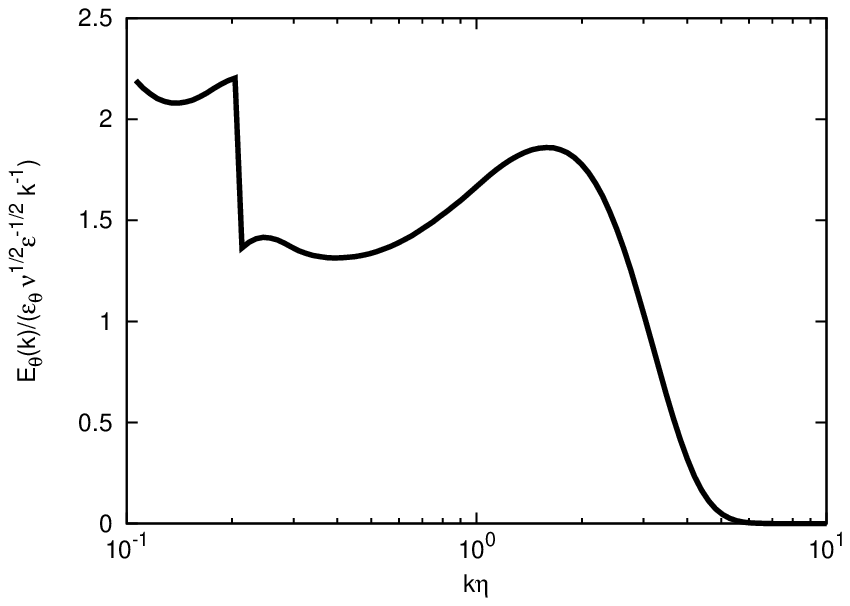}}
\caption{\label{FigETkSc} Passive scalar variance spectrum corresponding to a stationary passive scalar, injected in the large scales at small and high Schmidt numbers. Scalar spectrum at $Sc=0.01$ (a) and in compensated form (b). $Sc =20$ (c) and in compensated form (d). }
\end{center}
\end{figure}

\section{Numerical results\label{sec:Num}}

In this section we numerically integrate equations (\ref{eqF10}), (\ref{eqEK}) and (\ref{eqEKT}). 
%The latter equation being used only in the new model and not in the EDQNM studies. 
In these expressions we will insert for $\eta(k)$ and $\eta^{\theta}(k)$:

\begin{eqnarray}
\eta(k)=E(k,t)/F(k,t); \quad \eta^X(k)=0; \quad \eta^\theta(k)=\alpha k^2
\end{eqnarray}
%In all the other tests we will keep $\eta^\theta(k)=\alpha k^2$.}
%\item{The current model but as it was presented in (\cite{Bos2006}),
%\begin{eqnarray}
%\eta(k)=E(k,t)/F(k,t)+\nu k^2; \qquad \eta^X(k)=0; \qquad 
%\end{eqnarray}
%Indeed in that work not all arguments and assumptions were worked out and a viscous frequency was added to $\eta(k)$ with respect to the present work.}

We will first consider decaying turbulence, starting from an initial distribution 
\begin{equation}
E(k)=B k^4 e^{-2\left(k/k_L\right)^2}
\end{equation}
with $B$ chosen to normalize the energy to unity. $k_L=10$ and the first wavenumber of the domain is $k_0=1$. Wavenumber space is discretized using a geometrical discretization $k_i=k_0 r^{i-1}$. This possibility in two-point closures to use a geometrical discretization allows to attain much higher Reynolds numbers than in direct numerical simulations for a much lower cost. The discretization in the present computations is approximately $15$ wavenumbers per decade. The viscosity is equal to $\nu=8\cdot 10^{-8}$. From the initial condition we let the turbulence decay until it reaches a self-similar state in which $e/\epsilon\sim t$, with the kinetic energy and viscous dissipation respectively determined by
\begin{eqnarray}
e=\int E(k) dk,  \qquad \epsilon=2\nu \int k^2E(k)  dk.
\end{eqnarray}
All results are evaluated in this phase and the Reynolds number, defined as
\begin{equation}
R_\lambda=\sqrt{\frac{20}{3}\frac{e^2}{\nu \epsilon}}
\end{equation}
is at the time of evaluation $5000$. The scalar spectra are initialized to have the same shape as the kinetic energy. Mixing at three different Schmidt numbers, $Sc=\nu/\alpha$, is evaluated, $Sc=0.01,1,20$. For the computations at $Sc=0.01$ and $Sc=20$, in order to clearly produce scaling behavior, both the velocity and the scalar field are forced at the lowest wavenumbers, $k\leq 2$. For the case in which $Sc=20$, The Reynolds number is taken equal to $R_\lambda=100$ and the discretization is refined to $50$ wavenumbers per decade to more acurately capture nonlocal interactions which play an essential role in the physics of the Batchelor range.

\subsection{Results for the velocity field for high Reynolds number
  decaying turbulence}

Results for decaying isotropic turbulence at $R_\lambda=5000$ are
shown in Figure \ref{FigEk1}. We observe that the Kolmogorov constant is of order $1.7$. 
%For EDQNM this constant is of order $1.4$. 
It is observed that in the present model the bottleneck effect, that is, the increasing value of the compensated plot in the beginning of the dissipation range, is absent. Note that this bottleneck is far more pronounced in the EDQNM model \cite{Andre1977}. Indeed, the bottleneck is generated by the fact that nonlocal interactions are cut off in the dissipation range, thereby decreasing the transfer of energy so that it piles up \cite{Herring}. This effect is very sensitive to the relative value of the viscous time-scale compared to the local (in $k$-space) turbulent time-scale. For example, the bottleneck can be increased by decreasing the value of $\lambda$ in the EDQNM model.

The constant in
\begin{equation}
F(k)=C_F\epsilon^{1/3}k^{-7/3}
\end{equation}
is order $C_F\approx 2.5$ as shown in Figure \ref{FigEk2}. The energy transfer spectrum is shown in Figure \ref{FigEk3} (a). In Figure \ref{FigEk3} (b) we see that if we compare the
$\eta(k)$ given by our model to $\lambda\sqrt{\int_0^k s^2E(s)ds}$,
there is a short range in which the two are approximately
proportional, with a value of $\lambda\approx 0.5$ .

\subsection{Results for the mixing of an isotropic passive scalar}

In Figure \ref{FigETk} we show the spectrum of a passive scalar which is freely decaying. The initial spectrum is equal to the spectrum of the velocity field and the spectra are computed at the same time-instant as in Figure \ref{FigEk1}. A clear inertial-convective range is observed in which the scalar variance distribution is proportional to $k^{-5/3}$, in agreement with classical arguments \cite{Corrsin1951,Obukhov}. In the compensated representation it is observed that the dimensionless constant in 
\begin{equation}
E_\theta(k)=C_\theta \epsilon_\theta\epsilon^{-1/3}k^{-5/3}
\end{equation}
is around $0.3$ according to our closure. This is about one half of the generally observed value in experiments and numerical simulations \cite{Sreenivasan}. This underprediction is a well-known deficiency of Lagrangian History DIA and its relatives. It can be cured for by changing the representative on which the closure is based \cite{Kraichnan1978,Herring1979,Gotoh2000}. This would constitute an interesting perspective, but is not tried in the present work. We note that EDQNM predicts a more realistic value for $C_\theta$ \cite{Herring}. We stress, however, that this value is directly related to the parameter $\lambda'$. If this value was chosen equal to $\lambda$ in the cited reference, then the value for $C_\theta$ would be of the same order as the one predicted by our theory. We further observe that a large bottleneck is observed in all these results.

\subsection{Mixing of an isotropic passive scalar at small and large Schmidt number}

We also carried out computations for small and large Schmidt number and the results are shown in Figure \ref{FigETkSc}. It is observed in Fig. \ref{FigETkSc} (a) and (b) that the closure manages to reproduce the inertial diffusive range \cite{Batchelor1959-2},
\begin{equation}
E_\theta(k)=\frac{1}{3} C_k \epsilon_\theta \epsilon^{2/3}\alpha^{-3}k^{-17/3}.
\end{equation}
In Fig. \ref{FigETkSc} (c) and (d), we observe that also that the viscous-convective range is reproduced \cite{Batchelor1959},
\begin{equation}
E_\theta(k)=C_\theta' \epsilon_\theta \nu^{1/2}\epsilon^{-1/2}k^{-1}.
\end{equation}

\subsection{Influence of the correlation time-scale of the displacement vector}
\begin{figure}
\begin{center}
\setlength{\unitlength}{.5\textwidth}
\subfigure[]{\includegraphics[width=1\unitlength]{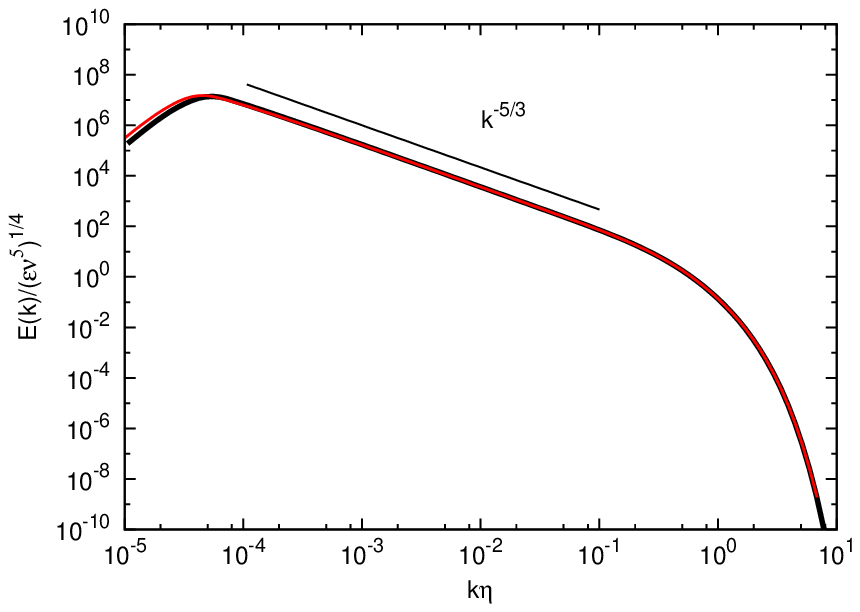}}~
\subfigure[]{\includegraphics[width=1\unitlength]{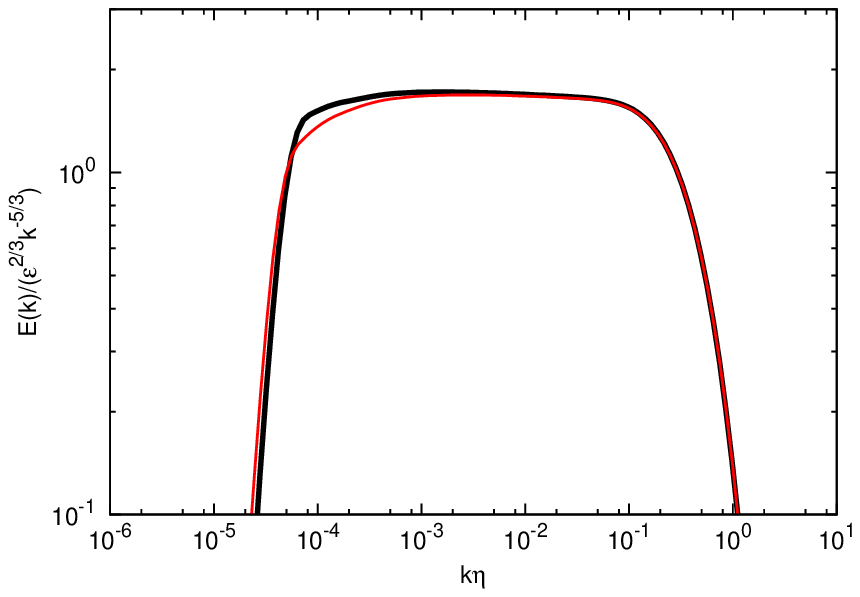}}
\caption{\label{Fig:TINT} Influence of the correlation time-scale of the displacement vector on the energy spectrum. Energy spectrum (a). Black corresponds to $\eta^X(k)=0$. The red graph corresponds to $\eta^X(k)=\mathcal{T}^{-1}$. Compensated spectrum (b).}
\end{center}
\end{figure}
To test the influence of the correlation time-scale of the displacement vector on the results, we performed a simulation, in which we changed the time-scale of the displacement vector from zero ($\eta^X(k)=0$) to the integral time-scale ($\eta^X(k)=\mathcal{T}^{-1}$), as described in the last paragraph of section \ref{Derivation}. The results are shown in Figure \ref{Fig:TINT}. Clearly, the influence of this change in correlation time is not dramatic in the current application. Only in the large scales of the energy spectrum a small influence is observed. In the inertial and dissipation range the change in time-scale does not seem to influence the results. The reason for this is that in $\Theta^F(kpq)$, only one of the three wave-vectors, $\bm k$ is influenced by this change. The  $\eta(p)$ and $\eta(q)$ rapidly dominate over the $\eta^X(k)$, since $\eta(k)$ is an increasing function of $k$ in the inertial range.

\section{Concluding remarks}

In the present work we derived a self-consistent Markovian closure starting from the DIA formalism and using the fluctuation-dissipation assumption. Closure was obtained by relating the turbulence time-scale to the spectrum of the cross-correlation of fluid-particle-displacement  and velocity, $F(k,t)$. The general idea was announced in a letter \cite{Bos2006}, but the details of the idea were not given there. {\bf In the present work we show how the present closure is related to the Abridged Lagrangian History DIA, and how the introduction of the displacement vector allows to measure the Lagrangian decorrelation in an Eulerian reference frame. We want to stress that such a single-time closure could perhaps also be obtained applying a similar procedure to the two-time correlations in the Lagrangian Renormalized Approximation, which was not tried here.}

%The present closure was however discovered, starting from EDQNM type procedures, whereas the Lagrangian Renormalized Approximation was directly obtained from a DIA procedure, applied to the velocity and position function. A similar procedure was presented in the present work.

The closure was numerically integrated at large Reynolds number ($R_\lambda=5000$). Results were presented for the energy spectrum and transfer spectrum. Results for the mixing of an isotropic passive scalar were shown for Schmidt numbers $0.01$, $1$ and $20$ respectively and classical scaling was reproduced.

In starting from DIA it becomes clear why a certain number of terms should be neglected in previous works. These terms were present in work on the velocity-scalar cross-correlation spectrum \cite{Herr,Ulitsky,Bos2005,Bos2007-1} and correspond in these studies to nonlinear transfer terms which were linearly dependent on the mean-scalar gradient. They should be removed if consistency is required with first-order renormalized perturbation theories such as DIA. They appear as terms which are second order in the formal expansion parameter in the DIA approximation. 

Furthermore, a simplified model for the scalar flux spectrum was proposed (\ref{eq:FEDQNM}), with a level of complexity comparable to the  standard EDQNM model for the energy spectrum. 

%Comparison of the present closure with EDQNM shows qualitative agreement. 
Let us finish by noting the following: during the Marseille congress 50 years ago, Kraichnan presented his DIA \cite{Kraichnan1962} to the scientific community. Many things have changed since then: computational power has increased enormously, making possible the direct numerical simulation of laboratory experiments, advanced experimental techniques allow to visualize almost every feature of a turbulent flow. However, the theoretical description of turbulence, derived from the Navier-Stokes equations, has still not advanced enormously beyond DIA related approaches. The physical insights obtained from these triadic closures are still an indispensable building block of our description of turbulent flows.

\section*{Acknowledgements}

Robert Rubinstein is acknowledged for many discussions on DIA and related issues. We are thankful to the organizers of the Turbulence Colloquium Marseille 2011, Marie Farge, Keith Moffatt and Kai Schneider for having organized a very stimulating scientific meeting. Two referees have significantly contributed to the improvement of the present work through their comments and suggestions.

\bibliographystyle{tJOT}
%\bibliography{/home/bos/PUBLI/biblio}

\end{document}